\documentclass[twocolumn, showpacs, english, preprintnumbers, amsmath, amssymb, superscriptaddress, aps]{revtex4-1}
\usepackage{latexsym}
\usepackage{epsfig,psfrag}
\usepackage{dcolumn}
\usepackage{bm}
\usepackage{graphicx}
\usepackage{color}
\usepackage{esint}
\usepackage{babel}
\usepackage{amsfonts}
\usepackage{enumerate}
\begin{document}
\title{Two-impurity helical Majorana problem} 
\author{Erik Eriksson} 
\affiliation{Institut f\"ur Theoretische Physik, 
Heinrich-Heine-Universit\"at, D-40225 D\"usseldorf, Germany}
\author{Alex Zazunov} 
\affiliation{Institut f\"ur Theoretische Physik, 
Heinrich-Heine-Universit\"at, D-40225 D\"usseldorf, Germany}
\author{Pasquale Sodano}
\affiliation{International Institute of Physics, Universidade 
Federal do Rio Grande do Norte, 59078-400 Natal-RN, Brazil}
\affiliation{Departamento de Fisica T{\'e}orica e Experimental, Universidade
Federal do Rio Grande do Norte, 59072-970 Natal-RN, Brazil }
\affiliation{INFN, Sezione di Perugia, Via A. Pascoli, 06123 Perugia, Italy}
\author{Reinhold Egger}
\affiliation{Institut f\"ur Theoretische Physik, 
Heinrich-Heine-Universit\"at, D-40225 D\"usseldorf, Germany}
\affiliation{International Institute of Physics, Universidade 
Federal do Rio Grande do Norte, 59078-400 Natal-RN, Brazil}
\date{\today}

\begin{abstract}
We predict experimentally accessible signatures for helical Majorana 
fermions in a topological superconductor by coupling to two quantum dots 
in the local moment regime (corresponding to spin-$1/2$ impurities).
Taking into account RKKY interactions mediated by bulk and edge modes,
where the latter cause a long-range antiferromagnetic Ising coupling,  
we formulate and solve the low-energy theory for this two-impurity
helical Majorana problem.  In particular, we show that 
the long-time spin dynamics after a magnetic field quench 
displays weakly damped oscillations with universal quality factor.
\end{abstract}

\pacs{74.78.-w, 73.21.-b, 74.40.-n} 

\maketitle

\section{Introduction}

Over the past few years, several groups have 
reported first experimental signatures for the elusive Majorana bound state 
in superconducting hybrid devices \cite{mbs1,mbs2,mbs3,mbs4,mbs5}.
Majorana bound states exist near the ends of
topologically nontrivial one-dimensional (1D) superconductors,
and many proposals have appeared on how to probe them
\cite{rev1,rev2,rev3,rev4,rev5,law,erik,new1,new2,new3,new4}.  
Likewise, the boundary of a 2D topological superconductor (TS) can
host propagating gapless Majorana fermion states \cite{rev1,fu1},
with different properties as compared to the mostly studied Majorana bound 
states.  Recent experiments on InAs/GaSb \cite{exp1,exp2}, HgTe/CdTe \cite{exp3},
or topological insulator \cite{exp4} heterostructures with
$s$-wave superconductors have indeed reported an edge-state dominated
Fraunhofer pattern of the Josephson current. 

These developments  highlight the urgent need for realistic proposals on how to 
prepare, manipulate, and detect 1D Majorana edge states.
We here consider the case of a time-reversal symmetric 2D TS 
with counterpropagating (right- and left-moving) Majorana edge 
states of opposite spin polarization \cite{rev1,fu1}.  
A major obstacle to the detection of these ``helical'' Majorana modes
arises from their charge neutrality. 
As a consequence, transport experiments are difficult,  
requiring one to study thermal transport or interferometric devices 
\cite{asano,interfero1,interfero2,interfero3,interfero4}.
Moreover, although a zero-bias tunneling conductance peak 
due to the Majorana edge is expected \cite{law}, 
 similar peaks are also caused by other mechanisms \cite{rev3}.  

\begin{figure}
\centering
\includegraphics[width=9cm]{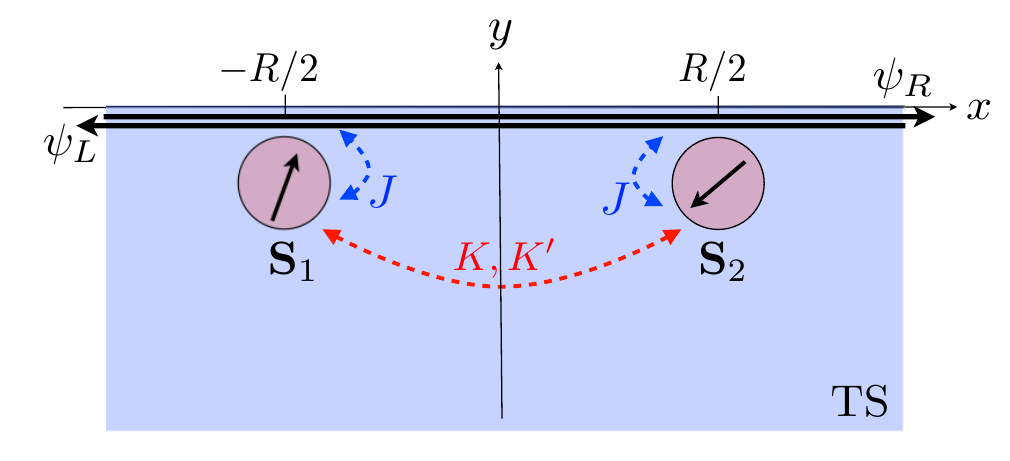}
\caption{\label{fig1} Schematic setup: Two quantum dots located
near the helical Majorana edge states of a 2D TS which occupies the $xy$
half-plane with $y<0$.  For strong on-site repulsion, the dots are 
equivalent to spin-$1/2$ operators, ${\bf S}_1$ and ${\bf S}_2$, respectively,
separated by a distance $R$.  They are exchange-coupled ($J$) 
to the local Ising spin density of the Majorana edge, and RKKY-coupled to 
each other ($K,K'$) through bulk TS modes.  
}
\end{figure}

We here propose a setup, see Fig.~\ref{fig1},
where the physics of helical Majorana fermions can 
be probed in a direct and realistic manner 
through their coupling to two quantum dots 
located near the boundary of a 2D TS, see Fig.~\ref{fig1}.
Assuming a standard parameter regime with strongly repulsive on-dot 
Coulomb interaction, each dot corresponds to a spin-$1/2$ operator \cite{altland}. 
We imagine that the dots are arranged slightly above the
superconducting plane, with an insulating layer separating them from the TS.
Tunnel couplings connecting the dots to the Majorana edge then 
imply the presence of an exchange coupling, $J$,
to the Ising spin density of the Majorana edge at the respective location.  (In what follows,
we often refer to the dots simply as ``spins.'')
The corresponding single-dot case was studied in Ref.~\cite{shindou}, and 
Kondo physics was found in a magnetic field.  However, the 
predicted quantum phase transition occurs only for unphysical 
parameter values \cite{simon}.  

For two dots, see Fig.~\ref{fig1}, we instead encounter a Majorana version of the 
classic two-impurity Kondo problem \cite{2impK}.  
In this setup, we predict clear signatures of the helical Majorana edge
to appear in the ``spin'' dynamics of the dots.  
The rich physics found for the two-impurity 
helical Majorana problem arises from the interplay between Ising-like
 exchange couplings and indirect Ruderman-Kittel-Kasuya-Yosida (RKKY) 
spin-spin interactions \cite{rkkygen} mediated by the TS. 
 We find that  the Majorana edge causes 
long-range RKKY contributions, which we nonperturbatively determine 
below. In addition, bulk TS modes mediate an SU$(2)$-symmetric ferromagnetic
RKKY coupling ($K$), plus an Ising-type antiferromagnetic 
coupling ($K'$) \cite{lossrkky}. These couplings are 
especially pronounced when the spins bind Shiba states inside the TS gap \cite{shiba}.

Before turning to derivations and detailed discussions, let us
briefly summarize our main results.  (i) The low-energy theory for the
two-impurity helical Majorana problem can be solved analytically.  
In the zero-field case, for a wide parameter regime,
we find that the ground state is a fully 
entangled triplet state of both spins even when they are widely
separated.  This
long-range entanglement is rather unique and of interest  in
quantum information applications \cite{horodecki}.  
(ii) The helical Majorana edge causes an
Ising-type antiferromagnetic RKKY coupling,
$K_M\sim J^2/R$, which exhibits a slow decay with 
spin-spin distance $R$. This result holds for arbitrary $J$, even though
conventional RKKY theory assumes small $J$.
(iii) Experimentally monitoring the weakly damped spin dynamics after a 
magnetic field quench allows one to extract clear signatures for 
helical Majorana fermions through the predicted universal quality factor,
see Eq.~\eqref{Qdef} below.  The magnetic fields can be produced by using
ferromagnetic finger gates.  Time-dependent measurements of the dot spins can be performed
with high precision using available state tomography techniques, 
see, e.g., Refs.~\cite{meas1,meas2,meas3}.  The setup in Fig.~\ref{fig1} thus 
offers a direct  
way to manipulate and detect the helical Majorana fermion edge state of a
2D time-reversal invariant TS. Apart from the above-mentioned platforms,
our predictions also apply to bilayer Rashba quantum wells \cite{nakosai}
and exotic triplet-paired superconductors \cite{rev1,rev2,rev3}.

The structure of the remainder of this paper is as follows. In Sec.~\ref{sec2},
we introduce the setup and the theoretical model to describe it.  We 
then turn to the perturbative regime of small $J$ in Sec.~\ref{sec3}, where we
derive the RKKY interaction mediated by the Majorana edge. In addition,
we provide an explicit solution of the problem at not
too low energy scales.  The low-energy regime is then discussed in Sec.~\ref{sec4},
where we show that the problem can be mapped to a dissipative quantum impurity
problem.  Concrete predictions for the resulting spin-boson like dynamics
are presented in Sec.~\ref{sec5}, where we also comment on the unique
signatures of helical Majorana fermions in such experiments.  Finally,
we conclude with a brief summary in Sec.~\ref{sec6}.
Throughout the paper, we work in units with $\hbar=k_B=1$.

\section{Model}
\label{sec2} 

We consider a time-reversal symmetric 2D TS located 
in the $xy$ half-plane $y<0$,
which hosts 1D helical Majorana fermions near the edge.
On energy scales below the bulk TS gap $\Delta$, 
the TS Hamiltonian reduces to the gapless edge contribution \cite{fu1}
\begin{equation}\label{h0}
H_0 = -iv \int_{-\infty}^\infty dx \left( \psi_R\partial_x \psi_R
-\psi_L \partial_x \psi_L \right),
\end{equation}
with self-adjoint Majorana field operators, $\psi_{\nu}^{}(x)=
\psi^\dagger_{\nu}(x)$, subject to the anticommutator algebra
\begin{equation}
\{\psi_\nu(x),\psi_{\nu'}(x')\}= \delta_{\nu\nu'}\delta(x-x'),
\end{equation}
where $\nu=R,L$. 
Weak time-reversal invariant perturbations, e.g., due to spin-orbit
coupling or elastic disorder, cannot gap out the edge state and may only 
renormalize the edge velocity $v$.

Importantly, because of the Majorana anticommutator algebra,
the spin density operator has only one nonvanishing component,
\begin{equation}\label{syx}
s(x)=i\psi_R(x)\psi_L(x).
\end{equation}
This Ising property reflects the Majorana ``half-fermion'' character
and is in contrast to helical Dirac fermions, where only
expectation values of the other spin density components vanish but
not the operators themselves.
The corresponding ``Ising direction'', $\hat e_I$, of the spin polarization 
depends on the actual TS realization \cite{rev1}.  
In most applications, the Ising direction is in the TS plane. We
therefore put 
\begin{equation}
\hat e_I=(\cos\theta, \sin\theta,0).
\end{equation}
For instance, using the same coordinates as in Fig.~\ref{fig1},
we find $\theta=0$ for the TS realizations in Refs.~\cite{shindou,lossrkky}, 
while $\theta=\pi/2$ for the Fu-Kane proposal \cite{fu1}.  

Next we consider two quantum dots at $x=\mp R/2$ near the edge of the TS. 
In experimental realizations, the dots could be
displaced away from the TS plane along the $z$-axis, with
finite tunnel couplings to both the Majorana edge
and to bulk TS modes, see Fig.~\ref{fig1}.
Taking dot parameters within the standard local-moment regime of strong 
on-site Coulomb interaction \cite{altland}, the dots 
correspond to spin-$1/2$ operators ${\bf S}_1$ and ${\bf S}_2$, 
respectively.  With $s(x)$ in Eq.~\eqref{syx},
tunnel couplings between the dots and the Majorana edge imply 
Ising exchange couplings \cite{simon},
\begin{equation}\label{hj}
H_J = J \left [ s(-R/2) S_{I,1}+ s(R/2) S_{I,2} \right],
\end{equation}
where we assume the same $J$ for both spins, and $S_{I,j=1,2}$ denotes the
projection of the respective spin operator to the Ising direction,
\begin{equation}\label{Ising}
S_{I,j} =\hat e_I \cdot {\bf S}_j = 
 S_{x,j}\cos\theta+ S_{y,j}\sin\theta.
\end{equation}

Although gapped bulk TS modes do not appear in Eq.~\eqref{h0}, 
they are important in mediating RKKY interactions between ${\bf S}_1$
and ${\bf S}_2$.  Treating the exchange coupling, $J_b$,
between the spins and the bulk TS modes perturbatively, 
see Ref.~\cite{lossrkky}, the full Hamiltonian reads 
\begin{equation}\label{fullmodel}
H=H_0+H_J+H_D,
\end{equation}
with 
\begin{equation}\label{hdot}
H_D = - K  {\bf S}_{1} \cdot {\bf S}_{2} + K' S_{y,1} S_{y,2} - 
{\bf B} \cdot ( {\bf S}_1+{\bf S}_2 ).
\end{equation}
Note that $H_D$ does not include the edge-mediated RKKY interaction.
For $R\ll \xi=v_b/\Delta$, where $\xi$ is the superconducting 
coherence length and $v_b$ the bulk Fermi velocity, the bulk-induced
RKKY couplings are estimated as 
\begin{equation}
K \approx \frac{J_b^2}{16\pi v_b R^3} ,\quad 
K'\approx \frac{3K}{2}.  
\end{equation}
For the case of arbitrary $R/\xi$, see Ref.~\cite{lossrkky}.
The SU$(2)$-invariant term $\sim K$ in Eq.~\eqref{hdot} 
favors ferromagnetic alignment of both spins, while the Ising
term $\sim K'$ favors antiferromagnetic alignment along the $y$-axis.

We note that similar RKKY interactions also appear in the conventional two-impurity
Kondo problem once spin-orbit couplings are taken into account 
\cite{mross,garst}.  However, here we have  $k_F=0$ because of particle-hole symmetry, 
and the usual $2k_F$-oscillations with distance, see Ref.~\cite{rkkygen}, are 
completely absent in our case. 

Finally, we have also included a magnetic Zeeman field proportional
to ${\bf B}=(B_x,B_y,B_z)$ in Eq.~\eqref{hdot}, 
which is assumed to act only on the total dot spin. 
Such a field could be produced by suitable ferromagnetic finger gates. 
Without loss of generality, we put $B_y\ge 0$. 

The model (\ref{fullmodel}) describes the two-impurity helical Majorana
problem. It differs from the two-impurity Kondo problem 
for Dirac fermions \cite{2impK}, which hosts Kondo physics and 
an unstable non-Fermi liquid fixed point.  This difference is 
present even when allowing for spin-orbit induced 
anisotropy effects in the conventional problem \cite{mross,garst}, and
can be rationalized by noting that for the Majorana case, we have
exchange couplings connecting the impurity spins to the Ising spin density 
only. As we show in detail below, this distinction leads to rather different physics 
as compared to the results in Refs.~\cite{2impK,mross,garst}.

\section{Majorana-mediated RKKY interaction and perturbative solution}
\label{sec3} 

We begin our analysis with the small-$J$ limit. The effect of the 
Majorana edge on the spin dynamics can then be taken into account by 
perturbation theory in $J$.  After some algebra, we obtain the additional 
Ising-like RKKY contribution,
\begin{equation}\label{rkkyeq}
H_M = K_M S_{I,1} S_{I,2},\quad
K_M =  \frac{ \pi v}{4R} (\rho_0 J)^2,
\end{equation}
with the density of states $\rho_0=1/(2\pi v)$  of the Majorana edge.
In contrast to the bulk-induced Ising-like RKKY term $\sim K'$ in 
Eq.~\eqref{hdot}, the anisotropy axis is now set by $\hat 
e_I$.  Note that the slow $1/R$ decay of $K_M$ is as for conventional 1D 
Dirac fermions. In fact,  by
replacing $\rho_0\to 2\rho_0$ in Eq.~(\ref{rkkyeq}), 
the $k_F=0$ Ising variant of the well-known 1D RKKY coupling 
\cite{yafet,dugaev} is recovered.  

For small $J$, the coupled spin dynamics is thus captured by the
effective Hamiltonian 
\begin{equation}
H_{\rm eff}= H_D+H_M.
\end{equation}  With
the total spin operator,
\begin{equation} 
{\bf S}={\bf S}_1+{\bf S}_2,
\end{equation} 
we then obtain
\begin{equation}\label{heff}
H_{\rm eff} = - \frac{K}{2} {\bf S}^2 + \frac{K_M\cos^2\theta}{2} S_x^2 
+ \frac{K'+K_M\sin^2\theta}{2}S_y^2 -{\bf B}\cdot {\bf S}.
\end{equation}
Using singlet ($S=M=0$) and triplet ($S=1$ with $M=0,\pm 1$) states
with spin quantization axis along the $y$-direction,
\begin{equation} \label{spinstates}
{\bf S}^2 |S,M\rangle = S(S+1) |S,M\rangle, \quad 
S_y|S,M\rangle = M|S,M\rangle,
\end{equation} 
the singlet state decouples and the remaining $3\times 3$ 
matrix representation for $H_{\rm eff}$ can be readily 
diagonalized.  

To illustrate the physics described by Eq.~\eqref{heff}, we now
consider the case $B_x=B_z=0$.  To simplify expressions, it is 
convenient to shift the overall energy scale such that
the singlet state $|0,0\rangle$, which is always an 
eigenstate, has the energy $E_{s}=K$. 
The triplet state $|1,0\rangle$, with $M=0$, is also an eigenstate with energy 
\begin{equation}
E_{t,0}= (K_M/2) \cos^2\theta.
\end{equation}
The $M=-1$ and $M=+1$ triplet states hybridize, resulting in the energies
\begin{equation}\label{spectrum}
 E_{t,\pm}= \frac{2K'+K_M(1+\sin^2\theta)}{4} 
\pm \sqrt{B_y^2+ (K_M/4)^2 \cos^4\theta}.
\end{equation}
We now discuss the resulting ground state for ${\bf B}=0$
with $\theta=\pi/2$ and $\theta=0$, respectively.

First, when the Ising direction is oriented along the $y$-axis, i.e.,
for $\theta=\pi/2$, 
noting that all RKKY couplings ($K,K',K_M$) are positive, we find
that the entangled triplet state $|1,0\rangle$ is
always the ground state.
This state minimizes both the ferromagnetic SU$(2)$-symmetric 
RKKY term $\sim K$ and the antiferromagnetic Ising RKKY 
interaction $\sim K'+K_M$.
For typical parameters \cite{lossrkky} and $R\approx 10$~nm, the 
excitation energies above this ground state correspond to 
temperatures $\approx 1\ldots 10$~K.  
At lower temperatures, both dots are therefore fully entangled 
due to TS-mediated long-range RKKY interactions.

Second, for $\theta=0$, using $K'\approx 3K/2$, we see in a similar
fashion that the ground state is either again the triplet state, $|1,0\rangle$ for $R>R_c$,
or it corresponds to $E_{t,-}$ for $R<R_c$.  As the spin-spin distance
$R$ is varied through a critical value $R_c$ determined by 
$K'(R_c)=K_M(R_c)$, we thus observe that a quantum phase
transition occurs. This transition is 
caused by the competition between the Majorana- and the bulk-induced 
Ising-like RKKY terms, which have different anisotropy axes 
for $\theta\ne \pi/2$.

\section{Low-energy theory}
\label{sec4} 

In this section, we turn to the low-energy regime and
thereby discuss the physics beyond the perturbative small-$J$ regime.
To start, let us combine the Majorana fields into a chiral Dirac fermion field,
see also Ref.~\cite{shindou},
\begin{equation}\label{chiralferm}
\Psi(x)= \frac{1}{\sqrt{2}} \left( \psi_R(x)+i\psi_L(-x)\right),
\end{equation}
where Eq.~\eqref{h0} yields the equivalent form
\begin{equation}
H_0  = -iv\int dx\ \Psi^\dagger \partial_x \Psi.
\end{equation}
The Ising exchange term in Eq.~(\ref{hj}) then reads
\begin{eqnarray}\label{hkdirac}
H_J & = & \frac12 JS_I \left[ \Psi^\dagger(R/2)\Psi(-R/2)+ {\rm h.c.} \right] 
\\ \nonumber &+& \frac12 J(S_{I,1}-S_{I,2}) 
\left[ \Psi^\dagger(R/2)\Psi^\dagger(-R/2)+ {\rm h.c.} \right] 
\end{eqnarray}
with the total spin 
\begin{equation}
S_I=S_{I,1}+S_{I,2}.
\end{equation}
Note that $H_J$ effectively describes non-local single-particle processes of either 
potential scattering or pairing type.  

On energy scales below $v/R$, however, the non-locality present in 
Eq.~\eqref{hkdirac} generates only corrections that are irrelevant in the 
renormalization group (RG) sense. Indeed,  we find that the low-energy expansion,
$H_J\to H_J^{(1)}+H_J^{(2)}$, comes from the local operators
\begin{eqnarray}\label{hkrg}
H_J^{(1)} &=& J  S_I \Psi^\dagger(0)\Psi(0) ,\\
H_J^{(2)} &=& \frac{JR}{2} 
(S_{I,1}-S_{I,2}) \Psi^\dagger(0)\partial_x\Psi^\dagger(0) + {\rm h.c.}\nonumber
\end{eqnarray}
The term $H_J^{(1)}$ is precisely marginal (scaling dimension 1), 
while the leading irrelevant operator $H_J^{(2)}$ has scaling dimension 2.
Performing a standard one-loop RG analysis, using the operator product expansions 
for the fermion operators in Eq.~\eqref{hkrg}, it immediately follows that the 
only new operator generated during the RG flow is precisely the RKKY term 
\eqref{rkkyeq}. The $H_J^{(1)}$ term does not renormalize, whereas the $H_J^{(2)}$ term flows 
to zero. Hence we conclude that $H_J^{(2)}$ can safely be taken into account by 
renormalization of $K_M$, and the fermionic low-energy theory is given by 
\begin{equation}
H_f=H_0+H_J^{(1)}.
\end{equation}

To proceed further, we now bosonize the chiral fermion in Eq.~\eqref{chiralferm},
see Ref.~\cite{altland},
\begin{equation}
\Psi(x)=\frac{1}{\sqrt{2\pi R}} e^{-i\phi(x)},
\end{equation} 
using the chiral boson field $\phi(x)$ with commutator
\begin{equation}
[\phi(x),\phi(x')]_-= i\pi {\rm sgn}(x-x'),
\end{equation}
where $R$ is taken as short-distance cutoff length.
The Euclidean action corresponding to $H_f$ is then given by
\begin{eqnarray}\nonumber
S_f &=& -\frac{1}{4\pi}\int dx d\tau \ 
\partial_x\phi \ (i\partial_\tau+v\partial_x)\phi \\ \label{bosact}
&+& v\rho_0 J\int d\tau S_I(\tau) \partial_x\phi(0,\tau),
\end{eqnarray}
where $S_I(\tau)$ is a discrete imaginary-time spin path.
The benefit of this step is that we have a Gaussian action for the bosonic 
field variables, which can therefore be integrated out exactly. 

Performing the Gaussian field integration over $\phi$, we finally
obtain the effective spin action 
\begin{eqnarray} \nonumber
S_{\rm spin} &=& -\frac12 (\rho_0 J)^2 \int d\tau d\tau'
\frac{S_I(\tau)  S_I(\tau')}{(\tau-\tau')^2+(R/v)^2} \\ 
&+&  (v/R) (\rho_0 J)^2 \int d\tau S_I^2(\tau).  \label{seff}
\end{eqnarray}
The first term corresponds to Ohmic damping \cite{shindou,weiss,foot1}.
We mention in passing that this term
vanishes for a constant spin path, $S_I(\tau)= 0,\pm 1$.
The second term instead describes once again the RKKY coupling $K_M$ 
due to the Majorana edge.  As expected from the chiral anomaly
of 1D fermions \cite{altland}, no orders higher than $J^2$ 
appear in Eq.~\eqref{seff}.  This indicates that up to a prefactor 
of order unity, Eq.~\eqref{rkkyeq} stays in fact valid beyond the 
perturbative small-$J$ regime.

We now show that, up to an overall irrelevant energy shift, the
low-energy Hamiltonian for the two spins can be written as 
\begin{equation} \label{finalH}
H = H_{\rm eff}+ \left(S_x \cos\theta+S_y\sin\theta
\right) {\cal E} + H_B [{\cal E}], 
\end{equation}
where ${\cal E}$ is a Gaussian random field with zero mean 
and $H_{\rm eff}$ is defined in Eq.~\eqref{heff}.
The ``bath'' Hamiltonian $H_B$ here describes an infinite set of 
harmonic oscillators generating the correlation function 
\begin{equation} \label{ldef1}
L(z)= \langle {\cal E}(0) {\cal E}(z) \rangle_B 
\end{equation}
for complex time $z=t-i\tau$.  To show the correctness of Eq.~\eqref{finalH}, 
we note that when averaging the partition function corresponding to Eq.~\eqref{finalH} 
over the Gaussian random field ${\cal E}$, we should arrive back at the 
effective two-impurity action (\ref{seff}) \cite{weiss}. 
This procedure allows us to determine the bath correlation function $L(z)$.
Indeed, 
by averaging over ${\cal E}$ and comparison to Eq.~\eqref{seff}, we find
that  Eq.~\eqref{ldef1} must be given by
\begin{equation}\label{ldef}
L(z) =  \frac{1}{\pi} \int_0^{\infty} d\omega J(\omega) 
\frac{\cosh[\omega (-iz+\beta/2)]}{\sinh[\beta\omega/2]}.
\end{equation}
In Eq.~\eqref{ldef}, we allow for finite temperature $T=1/\beta$ and
state the result for complex time $z=t-i\tau$.
The correlation function $L(z)$ is here expressed in terms of a so-called 
Ohmic spectral density \cite{weiss},
\begin{equation}\label{ohmic}
J(\omega) = 2\pi \alpha \omega e^{-\omega/(v/R)},\quad 
\end{equation} 
which in turn contains the dimensionless damping parameter
\begin{equation}\label{ohmic2}
\alpha= \frac12(\rho_0 J)^2.
\end{equation}
For physically relevant parameters, 
the damping strength is small, $\alpha\ll 1$.
For $T=0$, Eq.~\eqref{ldef} then yields a characteristic inverse-square
time dependence,
\begin{equation}
L_{T=0}(z)= - \frac{2\alpha}{(z-iR/v)^2}.
\end{equation} 
The general formulation in Eq.~\eqref{finalH} offers a convenient starting point 
to discuss the dissipative real-time dynamics of the two coupled spins.

\section{Dissipative two-impurity spin dynamics}
\label{sec5}

In what follows, we consider this effective spin dynamics in a
magnetic field and show that it contains unique signatures
of the underlying helical Majorana fermions.  For clarity, we here choose the   
Ising direction corresponding to $\theta=\pi/2$ and put $B_z=0$,
but we believe that our conclusions apply generally.

Using the parameter 
\begin{equation}
\epsilon=-B_y+ (K'+K_M)/2,
\end{equation} 
we first observe from Eq.~\eqref{spectrum} that for $B_x=0$,
the field component $B_y$ along the Ising direction
drives a quantum phase transition from the entangled $|1,0\rangle$ state, 
which is the ground state for $\epsilon>0$,
to the separable polarized triplet state $|1,1\rangle$ 
for $\epsilon<0$. 
In fact, assuming that the system parameters are in the regime
\begin{equation}
{\rm max}\{|\epsilon|,|B_{x,z}|,T\} \ll K,
\end{equation}
it is justified to project Eq.~\eqref{finalH} to the subspace spanned
by $|1,1\rangle$ and $|1,0\rangle$ only.
Defining Pauli matrices $\sigma_{x,z}$ in that subspace, 
such a projection arrives at the well-known spin-boson model \cite{leggett},
\begin{equation}\label{spinboson}
H_{\rm SB} = -\frac{B_x}{\sqrt{2}} \sigma_x+ 
\frac{\epsilon + {\cal E}}{2}\sigma_z + H_B.
\end{equation}

Since the damping parameter $\alpha$ in Eq.~(\ref{ohmic2}) is small,
the spin dynamics found after a sudden change (``quench'') of
the magnetic field will then show weakly damped oscillations.   
For instance, taking a constant field component $B_x$ and suddenly 
switching $B_y$ at time $t=0$, such that we have a 
large negative $\epsilon$ at $t<0$ but 
a vanishing value afterwards, $\epsilon(t>0)=0$,
the $T=0$ spin dynamics follows from the exact long-time result \cite{saleur}
\begin{equation}\label{dampedosc}
\langle \sigma_z(t)\rangle \sim e^{-\Gamma t} \cos(\Omega t), 
\end{equation}
where the average is taken using $H_{\rm SB}$.  
The physical spin dynamics then follows from the relation
\begin{equation}
\langle S_y(t)\rangle= \frac12\left(\langle \sigma_z(t)\rangle+1 \right).
\end{equation}
Equation (\ref{dampedosc}) is characterized by the damping rate
\begin{equation}\label{dampings}
 \Gamma= 2T_b \sin^2 \left(\frac{\pi \alpha}{2(1-\alpha)}\right),
\end{equation}
and by the quality factor $Q$
connecting the damping rate to the oscillation frequency,
\begin{equation}\label{Qdef}
Q= \frac{\Omega}{\Gamma} = \cot\left( \frac{\pi \alpha}{2(1-\alpha)}\right).
\end{equation}
Importantly, the quality factor is universal in the sense that it 
is independent of $B_x$ or other microscopic parameters.
In Eq.~\eqref{dampings}, we use the energy scale
\begin{equation}
T_b = c_b (R B_x/v)^{\alpha/(1-\alpha)} B_x,
\end{equation}
where $c_b$ is a prefactor of order unity (for its precise value, 
see Ref.~\cite{weiss}). 
The above reasoning suggests that an experimental observation of damped spin 
oscillations with universal ($B_x$-independent) quality factor, see Eq.~\eqref{Qdef}, 
constitutes a nontrivial signature for helical Majorana fermions.  

\section{Conclusions}\label{sec6}

In this work, we have proposed to couple two quantum 
dots in the local moment regime (where they correspond to spin-$1/2$ impurities)
to the helical Majorana edge states of a 2D topological superconductor.  
The superconductor then causes edge- and bulk-induced RKKY interactions 
among the two spins, where the 
resulting low-energy theory  can be solved in an essentially 
exact manner.  Notably, the physics found for this
two-impurity helical Majorana problem 
strongly differs from the respective
helical Dirac problem, see also Ref.~\cite{maciejko}. 
In the presence of time-dependent magnetic fields,
the spin dynamics is characterized by weakly damped 
coherent oscillations with universal quality factor. 
In a setup as shown in Fig.~\ref{fig1}, we believe that 
such oscillations could  be reliably measured by 
available state tomography techniques \cite{meas1,meas2,meas3}.

\acknowledgments

We thank A. Tavanfar for useful discussions. 
R.E.~acknowledges support by SPP 1666 of Deutsche Forschungsgemeinschaft
and by CNPq program Science Without Borders (SWB).
P.S.~thanks the Ministry of Science, Technology, and Innovation of Brazil 
and CNPq for granting a ``Bolsa de Produtividade em Pesquisa'',
and  acknowledges support by the CNPq SWB program,
and from MCTI and UFRN/MEC (Brazil).


\begin{thebibliography}{99}
\bibitem{foot1} Bulk TS modes are gapped and produce no low-frequency
contributions to damping.
\end{thebibliography}
\end{document}